\documentclass{article}
\usepackage{amsfonts}
\usepackage{amsmath,amsthm}
\usepackage{graphicx,epstopdf,epsfig,multirow,epic,bm}
\usepackage{color}
\usepackage{multicol}
\usepackage{CJK}

\normalsize \rm
\begin{document}
\begin{center}
{\huge \textbf{Handshake between Fibonacci series and pure preferential attachment
mechanism on a graph-model}}\\[12pt]
{\large Fei Ma$^{a,}$\footnote{~The author's E-mail: mafei123987@163.com. },\quad  Ding Wang$^{a,}$\footnote{~The author's E-mail: wangdingg@pku.edu.cn },\quad  Ping Wang$^{a,b,}$\footnote{~The corresponding author's E-mail: pwang@pku.edu.cn.} \quad  and  \quad Bing Yao$^{c,}$\footnote{~The corresponding author's E-mail: yybb918@163.com.}}\\[6pt]
{\footnotesize $^{a}$ School of Electronics Engineering and Computer Science, Peking University, Beijing 100871, China\\
$^{b}$ School of Software and Microelectronics, Peking University, Beijing  102600, China\\
$^{c}$ College of Mathematics and Statistics, Northwest Normal University, Lanzhou  730070, China}\\[12pt]
\end{center}

\begin{quote}
\textbf{Abstract:}  In order to better understand dynamical functions on amounts of natural and man-made complex systems, lots of researchers from a wide range of disciplines, covering statistic physics, mathematics, theoretical computer science, and so on, have spent much time in doing this intriguing study. In this paper, the discussed popularly topic, how to construct reasonable graph-model and then to explain many features of realistic networks using previously presented theoretical models, is still our main work. Compared with many pre-existing deterministic graph-model in single evolution way, our new graph-model can be constructed using three types of growth ways to meet preferential attachment mechanism. Meanwhile several typical indices associated with network research will be reported. In addition, some interesting findings will be shown, including the first handshake between Fibonacci series and ¡°pure¡± preferential attachment mechanism, an obvious relationship connecting two well-known rules, power-law and Zipf-law, and a common but useful equation on the basis of both spanning trees number and the number of spanning trees with maximum leaves. Based on these foregoing discussions, we can demonstrate that our graph-model obeys power-law and small-world property. For the future research directions, we present some unknown problems to be studied at the end of this paper.


\end{quote}

\vskip 1cm

\section{Background}

As an important and prominent member of a variety of complex system research, complex network study has taken much attention from different kinds of disciplines in the last few decades, such as statistic physics, mathematics, biology, computer science and chemistry, as well as other sociologies, and so on \cite{B.A-2001}-\cite{Albert-1999}. During these studies, a number of interesting outcomes have been displayed. Particularly, two of them are the discoveries of scale-free feature \cite{Albert-1999-1} and small-world character \cite{Watts-1998}. Such both groundbreaking findings in fact unveil two essential features in the evolution process of complex networks, and hence lead research up to a peak. After that, more and more researchers pursuit this hot study ¡°wave¡±. Nonetheless, there are still lots of structural properties on complex systems, including dynamical function, growth mechanisms and fundamental constructing ingredients, to name just a few, not transparent for us and even unknown.

Included examples as diverse as the world wide web, social network and protein interaction network of cell are well described as an abstract mathematical object, so-named graph, which contains some nodes and edges connecting many pairs of nodes \cite{S-C-C-2007}-\cite{R-M-K-1972}. Hereafter, for brevity, both concepts, network and graph, will be used indiscriminately. In other words, a network model $N(V,E)$ is the same as a graph-model $G(p,q)$, where $V$ and $E$ denote the set of vertices and edges of network respectively, corresponding to notations $|V|$ and $|E|$ are the order (the number of vertices) and size (edge number) of the network. As we all know, it is using both theoretical research and real-time data simulation that almost all current work are done. On the one hand, we can generate various types of models to probe those topological structure properties which are popular on real-world networks. On the other hand, there are various flaws in these generated theoretical models themselves, for example, the difficulties and inconvenience for capturing some exact characters of real-life networks by employing stochastic models, and the lack of accurate depiction of real-world networks based on deterministic models. The two phenomena exist in present study process and maybe continue in the next few decades, whereas some intriguing and significant features, as mentioned in the first paragraph, have been obtained and built by the preceding two classes of models. Currently, most of already presented deterministic graph-model are built on the basis of single growth manner. How to design deterministic graph-model possessing rich growth ways has attracted a little attention. Motivated by such a fact, our aim is to introduce new elements into deterministic graph-model group. One of necessary ways to accomplish this task is to establish graph-model in terms of mixed growth ways \cite{S-C-2017}. Indeed, in this paper, we employ various growth ways to propose a new graph-model which can be viewed as a research direction in the future to generalize deterministic graph-model depicting complex networks.

To smoothly show our results and reasonably organize this paper, let us here recall several concepts. A network is said to be \textbf{\emph{sparse}}, if its average degree is proportional to a small constant, that is the number of vertices is on order of magnitude with edge number in the growth process of network. And then, if its vertex degree distribution follows the well-known rule, \textbf{\emph{power-law}}, namely the ratio of the number of vertices with degree more than or equal to $k$ and order of the whole network satisfies the expression $P_{cum}(k)\sim k^{1-\gamma}$ ($2<\gamma<3$), the so-called \textbf{\emph{scale-free}} feature will be found over this model. After that, a model is \textbf{\emph{small-world}}, if the existence of higher clustering coefficient, smaller diameter ($D\sim \ln (|V|)$ or even $D\sim \ln(\ln (|V|)$)) and lower average path length ($APL$) can be shown. At present, most of models published in articles to depict varieties of complex networks in nature and daily life have scale-free feature and small-world property \cite{Y-L-2015}-\cite{A.L-2001}. Similarly, we will prove our graph-model to be scale-free and small-world. Except for that, our graph-model has the following figures. As known, the famous \textbf{\emph{zipf-law}}, due to George Kingsley Zipf, a Harvard linguistics professor, can be easily seen around our daily-life world. One of amounts of credible examples is the study about the frequency of use of the word in English text. Zipf found that the frequency of occurrence of the $r$th word is inversely proportional to its rank, namely meeting $f_{r}\sim r^{-b}$ with $b$ close to unity \cite{G-K-Z-1949}. It can be easily seen that the term of power-law is similar to the zipf-law, but there are also some subtle differences between them which has been directly or indirectly reported in many published literatures \cite{M-E-J-N-2005}. To the authors' knowledge, no rigorous proofs are made to show that there is or not direct connection between power-law and zipf-law. Here, it may be for the first time that there is such a transparent discussion about the two common laws on deterministic graph-model. And then, people who have obtained education in primary school can be familiar with \emph{\textbf{Fibonacci series}}. Numerous of work about Fibonacci series have already studied in the past few hundred years. Inspired by Fibonacci series, one topic of this paper is to build the first handshake between Fibonacci series and ¡°pure¡± preferential attachment mechanism which is referred to as absolutely preferential attachment mechanism \cite{Albert-1999-1}. This handshake is one of our innovation points. As a vital constant of graph-model, the spanning trees number always plays an important role not just on understanding some structural features, but also on determining some relevant dynamical properties, including network security, random walks and percolation, and so on \cite{X-W-J-2014}-\cite{F-M-20181}. Besides the importance of spanning trees number itself, another object corresponding to spanning trees has also gotten much focus from different kinds of research regions and can be described in terms of the below expression.  Given a connected graph $G(p,q)$, how to guarantee that the resulting subgraph $\mathcal{G}(p,q')$ with the same vertex set as $G(p,q)$ is connected and has maximum leaves by making use of as few edges from the edge set of $G(p,q)$ as possible is a formidable and challenged task of graph theory and theoretical computer science. Nonetheless, based on the growth procedure of our graph-model, the last task is to look for an equality linking \emph{\textbf{spanning trees number}} with \textbf{\emph{the number of maximum leaves spanning trees}} with respect to structural figures.

This paper begins with an introduction to some basic and helpful concepts and notations of network in Section 2 \cite{Bondy-1976}. Next, we will generate a type of new graph-model and then compute the order and size over this graph-model in Section 3. Section 4 includes some detailed discussions and explanations about 4.1) average degree, which says our graph-model is sparse, 4.2) degree distribution, which states our graph-model has scale-free feature, 4.3)-4.4) clustering coefficient, diameter and average path length, which together bring a proof to our graph-model being small-world. As a widely used formula of Brouwer's fixed point theorem, lemma 1 will be introduced in Section 5. The following section, Section 6, mainly devotes to enumerating spanning trees on our graph-model and to finding a relationship between spanning trees number and the number of spanning trees with maximum leaves, which provides our future research with useful tools and techniques. We close this paper using an elaborated conclusion, made up both our innovation points and overview about our coming tasks to be done in the days to come, in the last section.

\section{Notation and Terminology}

In this section, we mainly introduce some notations and terminologies used in the rest of this paper. Firstly, let us claim Fibonacci series, also vividly referred to as ¡°\textbf{\emph{Rabbit Sequence}}¡±, then state in turn other definitions, as follows

\textbf{Definition 1} A sequence consisting of positive integers $F(i)$ ($i=1, 2, 3, ...$) can be called Fibonacci series, if its first three terms obey $F(1)=1$, $F(2)=1$, $F(3)=2=F(1)+F(2)$ and the forward elements ($i\geq 4$) all satisfy an equality

\begin{equation}\label{eqa:mf-2-1}
F(i)=F(i-1)+F(i-2)
\end{equation}
sometime also defined as classical Fibonacci series. Motivated by the result due to famous mathematician Fibonacci, we here briefly list few generalized terms related to Fibonacci series. The classical Fibonacci series has \textbf{\emph{length}} $2$, denoted by symbol $\tau =2$, which is equal to the number of elements in the right-hand side of Eq.(\ref{eqa:mf-2-1}). With the original definition and length, we may generalize the classical Fibonacci series to obtain a more general expression, also thought of as a \emph{\textbf{$\tau$-term Fibonacci series}}, following the below character

$$\mathcal{F}(1)=a_{1}, \mathcal{F}(2)=a_{2}, \mathcal{F}(3)=a_{3}, ..., \mathcal{F}(i)=a_{i},...$$
and
$$\mathcal{F}(j)=\sum_{i=j-\tau}^{j-1}\mathcal{F}(i) \quad(j>\tau).$$

After reviewing lots of published articles consisting of deterministic network models, we obtain a few growth manners in which the initial model will continuously evolute over time. Three types of known growth manners will be utilized to create our graph-model, which are introduced in the next definition.

\textbf{Definition 2} Given an initial network model $N_{i}(V_{i},E_{i})$, at time $t=i+1$ $(t\geq0)$, adding a new vertex for each old edge belonging to the edge set $E_{i}$ of $N_{i}(V_{i},E_{i})$ and linking the new vertex to two endpoints of that old edge will produce the next network model $N_{i+1}(V_{i+1},E_{i+1})$. Such an operation is vividly called \textbf{\emph{triangle-growth function}}, as shown in Fig.1(a). If we say the triangle-growth function just depends on older edges, the next another operating way will only associate with older vertices, usually called \textbf{\emph{star-growth function}}, namely only linking $k$ new vertices to degree $k$ old vertex, as depicted in Fig.1(b). In addition, we can yet join a new parallel edge connecting two new vertices for every old edge of model $N_{i}(V_{i},E_{i})$ and link corresponding vertex-pair to construct the next model $N_{i+1}(V_{i+1},E_{i+1})$. The last operation is typically defined as the \textbf{\emph{rectangle-growth function}}, seeing Fig.1(c).

\begin{figure}
\centering
  \includegraphics[height=3cm]{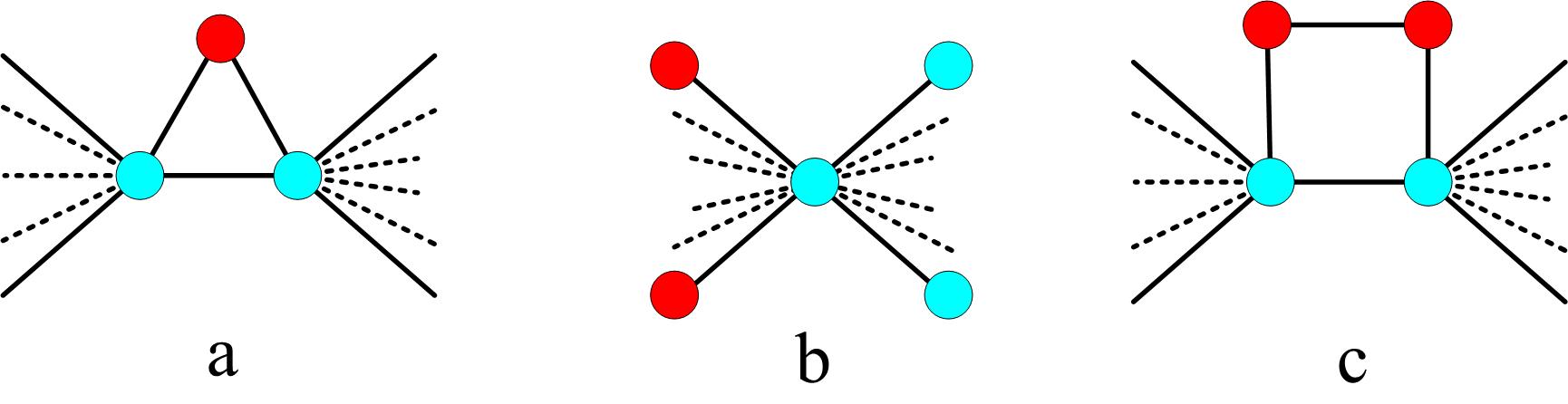}\\
{\small Fig.1. The diagram of growth operations. (a), (b) and (c) describes the \textbf{\emph{triangle-growth function}}, the \textbf{\emph{star-growth function}} and the \textbf{\emph{rectangle-growth function}} in which each indigo vertex and red represents old one and new, respectively.  }
\end{figure}

\textbf{Definition 3} For a given graph $N(V,E)$, a \textbf{\emph{spanning subgraph}} $N'(V,E')$ of graph $N(V,E)$ is a sub-graph with the same vertex set as $N(V,E)$ and a number of edges $E'$ such that $|E'|\leq|E|$. A \textbf{\emph{spanning tree}} $T(V,E')$ of a connected $N(V,E)$ is a spanning subgraph which is a tree having $|E'|=|V|-1$. If a spanning tree of a connected $N(V,E)$ has maximum leaves, then it is commonly called \textbf{\emph{maximum leaves spanning tree}}, shorted as \emph{\textbf{MLS-tree}}.

For brevity, table 1 summarizes some notations used later.

\small \textbf{Table 1.}
\begin{center}
\begin{tabular}{c|cc}
\hline
  $N(t)$&  The network model, at time step $t$  \\
\hline
$V(t)$&  The vertex set of network model $N(t)$ & \\
\hline
  $|V(t)|$&   The number of vertices of network model $N(t)$ (or the order of network model $N(t)$)&\\
\hline
$E(t)$&  The edge set of network model $N(t)$ & \\
\hline
  $|E(t)|$&   The number of edges of network model $N(t)$ (or the size of network model $N(t)$&\\
\hline
$\langle k \rangle$&   The average degree of network model;\qquad $\langle k \rangle$=$\frac{2|E(t)|}{|V(t)|}$&\\
\hline
$n_{k'}$&   The number of vertices with degree $k'$&\\
\hline
$P_{cum}(k)$&   The cumulative vertex-degree distribution of network model;\qquad $P_{cum}(k)$=$\frac{\sum_{k'\geq k}n_{k'}}{|V(t)|}$&\\
\hline
$e_{k}$&   The number of edges that actually exist between all $k$ nearest neighbors of degree $k$ vertex&\\
\hline
$c_{k}$&   The clustering coefficient of one vertex with degree $k$;\qquad $c_{k}$=$\frac{e_{k}}{k(k-1)/2}$&\\
\hline
$\overline{c}$&   The clustering coefficient of network model;\qquad $\overline{c}$=$\frac{\sum c_{k}}{|V(t)|}$&\\
\hline
$d_{uv}$&   The distance of a pair of vertices $u$ and $v$&\\
\hline
$D$&   The diameter of network model&\\
\hline
$\overline{d}$&   The average path length of network model, shorted as $APL$; \qquad $\overline{d}$=$\frac{\sum_{u,v\in N(t)}d_{uv}}{|V(t)|(|V(t)|-1)/2}$&\\
\hline
$\mathbb{S}$&   The number of spanning trees&\\
\hline
$\mathcal{S}$&   The number of maximum leaves spanning trees&\\
\hline
$\Psi$&   The leaf number of maximum leaves spanning tree&\\
\hline
\end{tabular}
\end{center}

\section{A new graph-model $N(t)$}

Here, we will propose a new graph-model $N(t)$ in more detail. Firstly, this new graph-model will grow up to our desired graph in a compound iteration manner which is in essence a hybrid product among triangle-growth function, star-growth function and rectangle-growth function. The first two steps are shown in Fig.2. Let us see such a growth process, as follow

\begin{figure}
\centering
  \includegraphics[height=7cm]{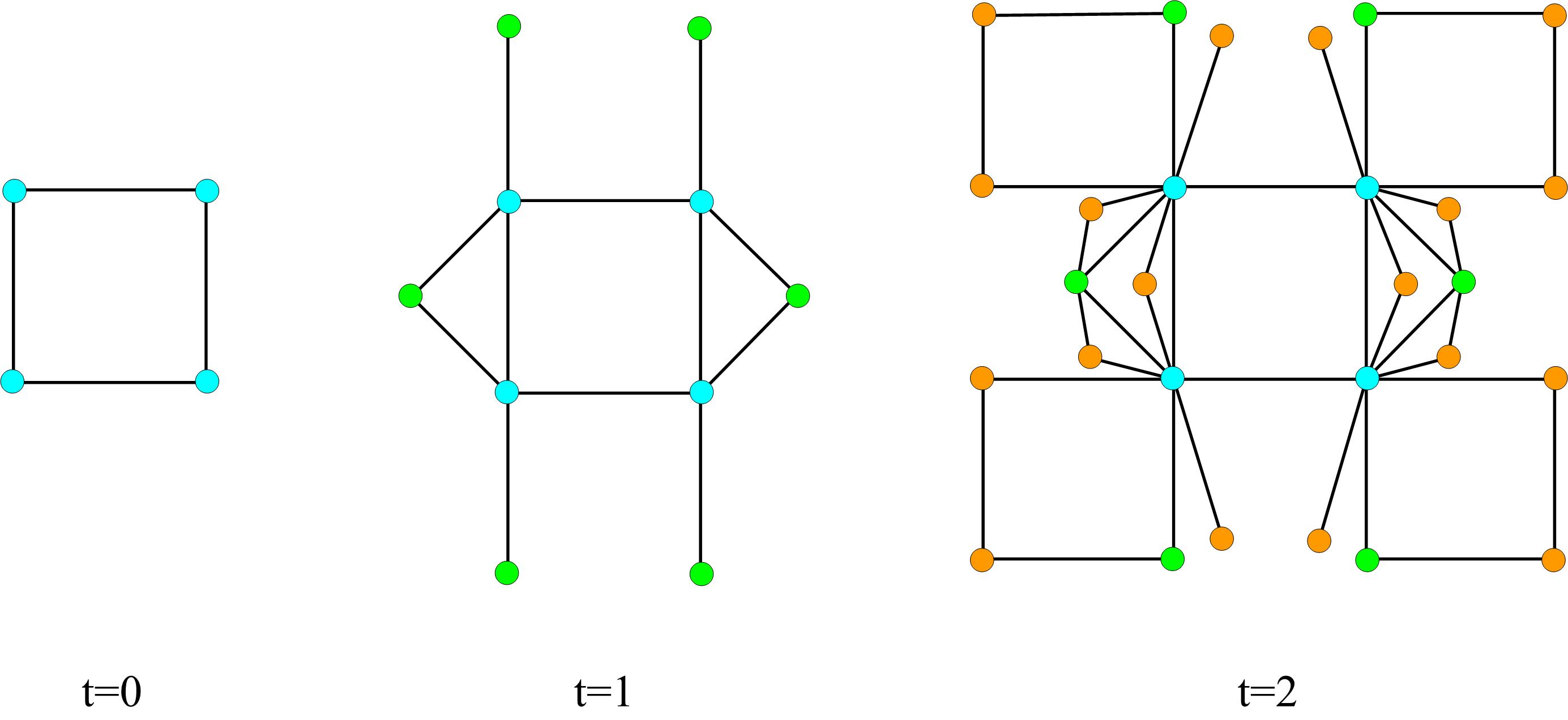}\\
{\small Fig.2. The diagram of the first two growth steps of graph-model $N(t)$.  }
\end{figure}

As $t=0$, the semina graph-model $N(0)$ is a rectangle that has four vertices, two vertical edges and another two aclinic.

As $t=1$, a newborn graph-model $N(1)$ is obtained from $N(0)$ by implementing two following operations

(1) For each of two vertical edges on $N(0)$, we apply triangle-growth function only one time;

(2) For each of two aclinic edges on $N(0)$,  we use star-growth function once exactly.

As $t=2$, except for adopting both triangle-growth function and star-growth function, we must make use of rectangle-growth function to build next graph-model $N(2)$. This process is described in much detail, as follows

(1) For each old edge of all triangles on $N(1)$, we just apply triangle-growth function only one time;

(2) For each old aclinic edge of all rectangles on $N(1)$, we use star-growth function once exactly;

(3) For each of all aclinic edges on $N(1)$ connecting a pendent vertex (also referred to as a leaf), we use rectangle-growth function once exactly;

As $t>2$, $N(t)$ is obtained from $N(t-1)$ by taking the above hybrid product mentioned at time step $t=2$. In order to fluently execute this growth procedure until our desirable model is built, we need make a statement (in fact without need, only to better describe) that at $t\geq2$, except for no any requirement for triangle-growth function, (a) all resulting rectangles must be arranged in a regular way, that is each generated rectangle should have two vertical edges and two aclinic, (b) all added edges using star-growth function must be vertical. If no, not only this produced model is formidable, but also the forward evolution process is obscure and hard to understand.

After $t$ time steps, our insight falls into the place to count vertex number $|V(t)|$ and $|E(t)|$ of $N(t)$. Obviously, the first several expressions can be easily gained by hand, such as $|V(0)|=4, |V(1)|=10, |V(2)|=28$ and $|V(3)|=78$, corresponding $|E(0)|=4, |E(1)|=12, |E(2)|=40$ and $|E(3)|=124$. It is not an effortless task to capture the exact solution to $|V(t)|$ and $|E(t)|$ in comparison with other published models, namely in general writing a group recursive equations among $|V(i)|$ and $|E(i)|$. With regards to the developing process of $N(t)$ ($t\geq3$), we may write

\begin{equation}\label{eqa:mf-3-1}
|V(t)|=L(t)+2L(t-1)+2(|V_{1}(t)|-2)+2\sum_{i=3}^{t}4^{i-2}(|V_{1}(t+1-i)|-2)+4^{t-1}
\end{equation}
and
\begin{equation}\label{eqa:mf-3-2}
|E(t)|=L(t)+4L(t-1)+2|E_{1}(t)|+2\sum_{i=3}^{t}4^{i-2}|E_{1}(t+1-i)|+\frac{1}{2}L(t-1)
\end{equation}
where symbol $L(t)$ represents those aclinic edge number added by star-growth function at time step $t$. For the explanation to symbols $|V_{1}(t)|$ and $|E_{1}(t)|$, we need introduce a prevail model $N_{1}(t)$ which is obtained from $N_{1}(t-1)$, however $N_{1}(t-1)$ from $N_{1}(t-2)$, ..., $N_{1}(1)$ from an initial model $N_{1}(0)$, which has two vertices together with an edge linking them, by making use of triangle-growth function for the unique old edge, hence continuously applying triangle-growth function with inverse order can rebuild model $N_{1}(t)$ after $t$ time steps. For the sake of the outline of paper, we omit the specific construction of $N_{1}(t)$. Readers can refer \cite{F-J-Y-B-G-2018} for further understanding. The $|V_{1}(t)|$ and $|E_{1}(t)|$ are vertex number and edge number on model $N_{1}(t)$ respectively. It is indeed easy to compute $|V_{1}(t)|$ and $|E_{1}(t)|$, as follows

\begin{equation}\label{eqa:mf-3-3}
\left\{\begin{split}&|V_{1}(t)|=|V_{1}(t-1)|+|E_{1}(t-1)|\\
&|E_{1}(t)|=|E_{1}(t-1)|+2|E_{1}(t-1)|
\end{split}
\right.
\end{equation}

Using the initial condition values $|V_{1}(1)|=3$ and $|E_{1}(t)|=3$ yields

\begin{equation}\label{eqa:mf-3-4}
|V_{1}(1)|=\frac{3^{t}+3}{2},\qquad|E_{1}(t)|=3^{t}.
\end{equation}

Plotting Eq.(\ref{eqa:mf-3-4}) and $L(t)=\frac{4^{t}-4}{3}$ (explained subsequently in much detail) into Eqs.(\ref{eqa:mf-3-1})-(\ref{eqa:mf-3-2}), we get

\begin{equation}\label{eqa:mf-3-5}
\left\{\begin{split}&|V(t)|=\frac{17\times4^{t-1}-3^{t}-11}{3}\\
&|E(t)|=\frac{28\times4^{t-1}-6\times3^{t-1}-22}{3}
\end{split}
\right. \qquad\qquad\qquad t\geq3
\end{equation}

It can be easily proved the values of $|V(3)|$ and $|E(3)|$ satisfy Eq.(\ref{eqa:mf-3-5}), respectively.

\section{Topological parameters}

As described in Section 1, we should start to analyze and discuss those main topological structure parameters over our graph-model $N(t)$ after achieving the construction of $N(t)$, such as 1) average degree, 2) degree distribution, 3) clustering coefficient, 4) diameter and average path length.

\subsection{Average degree}

As an index showing whether a given graph-model is sparse or not, the value of average degree can be relatively easily obtained using directly $|V(t)|$ and $|E(t)|$.

\textbf{Proposition 1} The solution of average degree $\langle k \rangle$ of graph-model $N(t)$ is

\begin{equation}\label{eqa:mf-4-1-1}
\langle k \rangle=\frac{2|E(t)|}{|V(t)|}\approx3.294.
\end{equation}

Thank to the smaller average degree value, our graph-model $N(t)$ is sparse according to \cite{Charo-2011}.

\subsection{Degree distribution}

As mentioned in Introduction, an interesting character over complex networks, the scale-free feature, is covered due to Barab\'{a}si and Albert by taking useful advantage of degree distribution \cite{Albert-1999}. Since, the degree distribution is defined as a standard verifying whether or not a given network model is scale-free. Before continuing this process, we need to make an explicit explanation for the ¡°pure¡± preferential attachment mechanism included in title. In the first research paper about complex network authored by both Barab\'{a}si and Albert \cite{Albert-1999-1}, they assumed there are two mechanisms, growth and preferential attachment, controlling the direction of network from disorder to obeying power-law, also called scale-free. A plausible stochastic model based on both growth and preferential attachment was built and its degree distribution was studied with the mean-field theory, in which the probability of any older degree $k$ vertex gaining new edges from new joining vertices absolutely relies on $k/\sum_{v\in N }k_{v}$. Subsequently, a series of derivative products correlated to preferential attachment claimed by Barab\'{a}si and Albert, that is to say, various types of preferential attachment manners, are successively brought to clarify the inner evolution mechanism over complex network as far as possible. Here, to distinguish between the former preferential attachment and the later, we add an adjective ¡°pure¡± at the head of the former. Now, it is time for us to determine whether or not graph-model $N(t)$ is scale-free.

\textbf{Proposition 2} The solution of degree distribution of graph-model $N(t)$ obeys

\begin{equation}\label{eqa:mf-4-2-1}
P_{cum}(k\geq k_{i})\propto k_{i}^{-\frac{\ln 4}{\ln 2}}=k_{i}^{1-\gamma},\quad \gamma=3
\end{equation}
where the power exponent $\gamma$ is equal to that obtained by mean-field theory \cite{Albert-1999-1}.

\textbf{Proof} In order to group different vertices of different degrees, we make the table 2 consisting of all vertices in decrease order of degree. Compared to the previous research \cite{Zhang-2014}-\cite{L-R-S-2017}, \cite{F-J-Y-2018},\cite{F-J-Y-B-G-2018}, table 2 includes an additional column that is a rank of all vertices with respect to degree, which helps us accomplish an intriguing finding.

\small \textbf{Table 2.}
\begin{center}
\begin{tabular}{c|c|ccccc}
\hline
   $rank$& $k$& $n$ \\
\hline
  $1$& $2^{t+1}$& $4$ \\
\hline
 $2$& $2^{t}$& $2$\\
\hline
 $3$& $2^{t-1}$& $18$ \\
\hline
 $4$& $2^{t-2}$& $38(=|V(3)|-|V(2)|-L(3)+2L(2))$ \\
\hline
  $...$& $...$& $...$\\
\hline
 $t-t_{i}+2$&   $2^{t_{i}}$& $|V(t+1-t_{i})|-|V(t-t_{i})|-L(t+1-t_{i})+2L(t-t_{i})$\\
\hline
  $...$&  $...$& $...$& \\
\hline
 $t+1$&  $2$& $|V(t)|-|V(t-1)|-L(t)+2L(t-1)$\\
\hline
 $t+2$&  $1$& $L(t)$\\
\hline
\end{tabular}
\end{center}

Because of other degree values not listed at table 2 are absent, it is necessary for obtaining the expression of degree distribution to adopt the cumulative method described in \cite{S.N. D-2002}. And then, we get

\begin{equation}\label{eqa:mf-4-2-2}
P_{cum}(k\geq k_{i})=\frac{\sum_{k\geq k_{i}}n_{k}}{|V(t)|}=\frac{24+\sum_{j=3}^{i+1}[|V(j)|-|V(j-1)|-L(j)+2L(j-1)]}{|V(t)|}\propto 4^{1-i}.
\end{equation}

Substituting $i=\frac{\ln k_{i}}{\ln 2}$ into Eq.(\ref{eqa:mf-4-2-2}), we obtain the desirable expression in Eq.(\ref{eqa:mf-4-2-1}). The proof is complete.

With the result in Eq.(\ref{eqa:mf-4-2-1}), we can report that our graph-model $N(t)$ follows power-law and has scale-free feature. After obtaining such an unexpected accomplishment, our work does not stop and still continues in its own way. Inspired by the pathbreaking work of George Kingsley Zipf, we study the problem between the frequency of occurrence of the $r$th vertex and its corresponding rank $r$ over graph-model $N(t)$, as follows

\begin{equation}\label{eqa:mf-4-2-3}
f_{r_{i}}=\frac{\sum_{r\geq r_{i}}n_{r}}{2|E(t)|}=\frac{4\times2^{t+1}+2\times2^{t}+18\times2^{t-1}+\sum_{j=4}^{r_{i}}2^{t-j+2}[|V(j-1)|-|V(j-2)|-L(j-1)+2L(j-2)]}{2|E(t)|}
\propto4^{1-i}.
\end{equation}

Combining Eq.(\ref{eqa:mf-4-2-3}) and the condition degree value $k_{r_{i}}=2^{t+2-r_{i}}$ of the $r_{i}$th vertex together produces

\begin{equation}\label{eqa:mf-4-2-4}
f_{r_{i}}\propto P_{cum}(k\geq k_{i}).
\end{equation}

Though there have been a large number of published papers all doing one identical thing to find a reasonable and acceptable bridge connecting power-law and zipf-law, it seems to be so intractable that a trivial viewpoint, ``they are different ways of looking at the same thing" \cite{L-A-A-2002}, has to be provided to make a helpless explanation. Now, we may here look at a light into the left problem and introduce an evidence connecting power-law and zipf-law over model $N(t)$. Maybe we can directly question all these deterministic scale-free models meet the below Eq.(\ref{eqa:mf-4-2-5})

\begin{equation}\label{eqa:mf-4-2-5}
\frac{f^{\lambda}_{r_{i}}}{P_{cum}(k\geq k_{r_{i}})}\propto O(1)
\end{equation}
and even make a conjecture almost all scale-free networks including natural and man-made obey  Eq.(\ref{eqa:mf-4-2-5}), in which exponent $\lambda$ is a constant as $t\rightarrow \infty$.

From the appearance of Eq.(\ref{eqa:mf-4-2-5}), our graph-model l in fact has $\lambda \approx 1$, a constant. We insist Eq.(\ref{eqa:mf-4-2-5}) will become a focus and a future research direction for physicians, mathematicians and linguists, as well other interested person. Particularly, it is said that zipf-law is popular in the distribution of passwords \cite{D-W-2017}. The importance of password is self-evident. Considering the complicate structure of widely used passwords, no fundamental theoretical model is reported to study the dynamical evolution on passwords, there are amounts of research results only based on simulation by sampling real-data on computer. It is possible for building basic theoretical model about password development to apply the result from Eq.(\ref{eqa:mf-4-2-5}).

Degree distribution is one of important topological parameters for probing complex network. After proving graph-model $N(t)$ to be scale-free, our main discussions will be diverted to answer whether or not graph-model $N(t)$ is small-world, namely whether graph-model $N(t)$ has higher clustering coefficient, smaller diameter and lower $APL$ or not.

\subsection{Clustering coefficient}

Clustering coefficient is referred to as another vital parameter of complex network research. Let us first recall the definition of clustering coefficient of a degree $k$ vertex, which is the ratio of the total number of triangles $e_{k}$ linking this vertex and all possible edge number among $k$ neighbor set of this vertex, $c=\frac{e_{k}}{k(k-1)/2}$. Now, we are at the place to calculate the solution of clustering coefficient of every vertex and the whole network model, respectively. To better organize this subsection, we need divide it into two smallest portions. For the first time, we have to list a table, as follows

\small \textbf{Table 3.}
\begin{center}
\begin{tabular}{c|c|c|ccc}
\hline
  $t$&  $\Delta(t)$& $\Theta(t)$& $L(t)$  \\
\hline
$1$&   $2$& $0$& $4$  \\
\hline
  $2$&   $0$& $4$& $4$\\
\hline
$3$&  $8(=2L(1))$& $4(=L(1))$& $4+4^{2}$ \\
\hline
  $4$&   $8(=2L(2))$& $12(=L(2))$& $4+4^{2}+4^{3}$\\
\hline
$...$&  $...$& $...$& $...$\\
\hline
$t_{i}$&  $2\times\frac{4^{t_{i}-1}-4}{3}(=2L(t_{i}-1))$& $\frac{4^{t_{i}-1}-4}{3}(=L(t_{i}-1))$& $\frac{4^{t_{i}}-4}{3}(=\Sigma_{i=1}^{t_{i}-1}4^{i})$\\
\hline
$...$&  $...$& $...$& $...$\\
\hline
$t$&  $2\times\frac{4^{t-1}-4}{3}(=2L(t-1))$& $\frac{4^{t-1}-4}{3}(=L(t-1))$& $\frac{4^{t}-4}{3}(=\Sigma_{i=1}^{t-1}4^{i})$\\
\hline
\end{tabular}
\end{center}
in which symbols $\Delta(t)$, $\Theta(t)$ and $L(t)$ are the incremental quantity of number of triangles, rectangles and vertical edges on $N(t)$, respectively. Then, the coming table 4 lists the main results on model $N_{1}(t)$, which is helpful to provide a solid fundamental explanation for clustering coefficient of our graph-model $N(t)$.

\small \textbf{Table 4.}
\begin{center}
\begin{tabular}{c|c|c|cccc}
\hline
   $n$& $k$& $\eta$& $c$\\
\hline
  $3^{0}$& $2^{t}$& $2^{t}-1$& $2^{-(t-1)}(=2\frac{2^{t}-1}{2^{t}(2^{t}-1)})$\\
\hline
 $3^{1}$& $2^{t-1}$& $2^{t-1}-1$& $2^{-(t-2)}(=2\frac{2^{t-1}-1}{2^{t-1}(2^{t-1}-1)})$ \\
\hline
 $3^{2}$& $2^{t-2}$& $2^{t-2}-1$& $2^{-(t-3)}(=2\frac{2^{t-2}-1}{2^{t-2}(2^{t-2}-1)})$ \\
\hline
  $...$& $...$& $...$& $...$\\
\hline
 $3^{i}$&   $2^{t-i}$& $2^{t-i}-1$& $2^{-(t-1-i)}(=2\frac{2^{t-i}-1}{2^{t-i}(2^{t-i}-1)})$\\
\hline
  $...$&  $...$& $...$& $...$ \\
\hline
 $3^{t-1}$&  $2$& $2^{1}-1$& $2^{0}(=2\frac{2^{1}-1}{2^{1}(2^{1}-1)})$\\
\hline
\end{tabular}
\end{center}
where symbol $\eta$ represents the number of triangles on $N_{1}(t)$.

Based on the above table 4, we can write the sum expression for the clustering coefficients of all vertices jointed by triangle-growth function into $N(t)$

\begin{equation}\label{eqa:mf-4-3-1}
\begin{split}\mathcal{C}_{1}=&\Delta(1)\sum_{i=0}^{t-1}3^{i}2^{-(t-1-i)}+\sum_{j=3}^{t}\Delta(j)\sum_{i=0}^{t-j}3^{i}2^{-(t-j-i)}
\end{split}
\end{equation}
where $\Delta(t)$ is defined at table 3.

Contrasted to capturing the result of Eq.(\ref{eqa:mf-4-3-1}), it is tricky to calculate the exact solution to those vertices added by another two operations, star-growth function and triangle-growth function. Before beginning this task, let us summarize a table at which an accidental phenomena is found

\small \textbf{Table 5.}
\begin{center}
\begin{tabular}{c|c|ccccc}
\hline
   $t$& $\alpha(t)$& $\beta(t)$\\
\hline
  $1$& $1$& $1(\textcolor[rgb]{1.00,0.00,0.00}{0})$ \\
\hline
 $2$& $1$& $1$ \\
\hline
 $3$& $2$& $1$ \\
\hline
  $4$& $3$& $2$\\
\hline
  $5$& $5$& $3$\\
\hline
  $...$& $...$& $...$\\
\hline
 $i$&   $\sum_{j=1}^{i-1}\beta(j)$& $\alpha(i-1)$\\
\hline
  $...$&  $...$& $...$\\
\hline
 $t$&   $\sum_{j=1}^{t-1}\beta(j)$& $\alpha(t-1)$\\
\hline
 $\textbf{\emph{Fibonacci series}}$ $(t \geq3)$&  $\alpha(t)=\alpha(t-1)+\alpha(t-2)$& $\beta(t)=\beta(t-1)+\beta(t-2)$\\
\hline
\end{tabular}
\end{center}
where symbols $\alpha(t)$ and $\beta(t)$ denote the incremental quantity of number of those edges added by star-growth function and that resulting rectangles built by rectangle-growth function on a vertex located at one having rectangle, respectively. If we replace the truth number value $\beta(1)=0$ (in red) with unity, it can been obviously understood that the Fibonacci series emerges suddenly at table 5, following

\begin{equation}\label{eqa:mf-4-3-2}
\alpha(t)=\alpha(t-1)+\alpha(t-2), \qquad \beta(t)=\beta(t-1)+\beta(t-2)
\end{equation}

Here provides a brief proof for reading

\begin{equation}\label{eqa:mf-4-3-3}
\left\{\begin{split}&\alpha(t)=\sum_{i=1}^{t-1}\beta(i)\\
&\alpha(t-1)=\sum_{i=1}^{t-2}\beta(i)\\
&\beta(t-1)=\alpha(t-2)\\
\end{split}
\right.
\end{equation}
together with

\begin{equation}\label{eqa:mf-4-3-4}
\left\{\begin{split}&\beta(t)=\sum_{i=1}^{t-2}\beta(i)\\
&\beta(t-1)=\sum_{i=1}^{t-3}\beta(i)\\
\end{split}
\right.
\end{equation}

Fibonacci series exists the whole development process of our graph-model $N(t)$. As the second innovation point of this paper, the handshake between Fibonacci series and ``pure" preferential attachment mechanism will be discussed in much detail at the end of Section 4 to close this section.

With the preceding statements, the sum of clustering coefficient of each vertex located at arbitrary rectangle on $N(t)$ can be obtained

\begin{equation}\label{eqa:mf-4-3-5}
\begin{split}\mathcal{C}_{2}=&\frac{4\times(\sum_{i=0}^{t-1}2^{i}+\sum_{i=1}^{t-2}\alpha(i)\sum_{j=i-1}^{t-3}2^{j})}{2^{t}(2^{t+1}-1)}+
\frac{4\times3\times(\sum_{i=0}^{t-3}2^{i}+\sum_{i=1}^{t-4}\alpha(i)\sum_{j=i-1}^{t-5}2^{j})}{2^{t-1}(2^{t}-1)}\\
&+\sum_{s=2}^{t-4}\frac{(4^{s}-4)(\sum_{i=0}^{t-s-2}2^{i}+\sum_{i=1}^{t-s-3}\alpha(i)\sum_{j=i-1}^{t-s-4}2^{j})}{2^{t-s-1}(2^{t-s}-1)}
+\frac{3(4^{t-3}-4)}{2^{2}(2^{3}-1)}+\frac{4^{t-2}-4}{2^{1}(2^{2}-1)}
\end{split}
\end{equation}

On the basis of both Eq.(\ref{eqa:mf-4-3-1}) and Eq.(\ref{eqa:mf-4-3-5}), we can have the proposition 3.

\textbf{Proposition 3} The solution of clustering coefficient $\overline{c}$ of graph-model $N(t)$ is

\begin{equation}\label{eqa:mf-4-3-6}
\overline{c}=\frac{\mathcal{C}_{1}+\mathcal{C}_{2}}{|V(t)|}\propto O(1).
\end{equation}

As the iteration step $t$ has a tendency to infinity, the average clustering coefficient $\overline{c}$ does not diverge but converge into an invariable number value independent of time step $t$, showing which our model has higher clustering coefficient.

\subsection{Diameter and average path length}

Subsection 4.3 has reported one characteristic index of small-world network, the two remained exponents will be studied at this place and the corresponding proposition may be gained. It is also beneficial to recall the definition of diameter $D$ and average path length $APL$. The diameter is the longest shortest path between all pairs of vertices, which portrays the maximum communication delay in the evolution of networks. Then the average path length can be defined as the ratio of the sum of distance of all pairs of vertices and the value of $|V(t)|(|V(t)|-1)/2$. For a given real-life growth complex network, it in general is formidable enough to capture the analytical expression of $APL$. Even for a deterministic one with regular topological structure, this sophisticated computing procedure always make researchers uncomfortable. We here take useful advantage of both topological structure of $N(t)$ and some techniques to attain our anticipant result and to avoid the headache calculations.

\textbf{Proposition 4} The solution of diameter $D(t)$ of graph-model $N(t)$ is

\begin{equation}\label{eqa:mf-4-4-1}
D(t)=2t+2\propto O(\ln|V(t)|).
\end{equation}

\textbf{Proof} The first three terms can be counted by hand, such as $D(0)=2, D(1)=4, D(2)=1+D(0)+1=4$. The following description will be processed by induction on iteration step $t$. Assume that this holds on step $i$, for time $i+1$, we may illustrate the recursive equation $D(i+1)=1+D(i)+1=2(i+1)+2$ by the three graphs shown in Fig.1.

\textbf{Case 1} For the leftmost one of Fig.1, adding a new vertex into $N(i)$ by triangle-growth function makes distance at most plus unity.

\textbf{Case 2} For the Fig.1(b), joining a new vertex into $N(i)$ by star-growth function makes distance absolutely add unity.

\textbf{Case 3} Similar to case 2, for the rightmost one of Fig.1, linking two endpoints of a new aclinic edge with two corresponding endpoints of an older aclinic edge in $N(i)$ by rectangle-growth function makes distance absolutely add unity.

We immediately get $D(i+1)=1+D(i)+1=2(i+1)+2$ using both the symmetrical structure of model $N(t)$ and case 1 - case 3. The proof is complete.

Given a relatively small graph, it is also an unmanageable matter to analytically calculate a precise expression of its $APL$ in comparison to capturing diameter $D$. Our graph-model $N(t)$ can sever as an example to verify the above demonstration, which is not our topic here. On the one hand, Dividing all vertices into different groups is hard such that the later process will have to be stop. On the other hand, enormous combination computation yet need to spend much time and effort. Apparently, it can be noticed that the following proposition does not report a concise number value for $APL$ but provide an approximate solution. An essential reason is that we just concentrate on how to reveal model $N(t)$ displays lower average path length and to further answer it has small-world property.

\textbf{Proposition 5} The solution of average path length $APL$ of graph-model $N(t)$ is

\begin{equation}\label{eqa:mf-4-4-2}
\overline{d}=\frac{\sum_{u,v\in N(t)}d_{uv}}{|V(t)|(|V(t)|-1)/2}\propto O(t).
\end{equation}

\textbf{Proof}. It can be simply proven by assertion on diameter $D$ of proposition 4. For more details see references \cite{F-J-Y-2018,Z-Z-Z-2011},\cite{F-J-Y-B-G-2018},\cite{ F-M-B-Y-2018}. It can be seen that $N(t)$ possesses small-world property.

By far, two most important characteristic indices, both scale-free feature and small-world property, have been indicated.
Now, let us put our eyes into studying the connection between Fibonacci series and ``pure" preferential attachment mechanism. Firstly, permit us to list some simple but useful real-life applications to Fibonacci series. (a) In cryptography. Based on Fibonacci series self-similarity and the sensitivity of logistic mapping to the initial value under chaotic condition, taking useful advantage of both Fibonacci series and chaotic mapping can create a chaotic pseudo-random sequence for cryptograph to protect ciphertext from attack \cite{T-Q-Z-2018}. (b) In geosciences. Authors \cite{S-B-2018} propose to apply Fibonacci lattices first as a tool to evaluate map projection distortions and then to optimize their defining parameters so that the resulting map projection has minimum distortion for a particular area of use, to name but a few \cite{W-Y-L-2018,D-l-A-2016}. Next, no surprising, this degree exponent of our graph-model $N(t)$ is the same as BA model's \cite{Albert-1999}. To some extent, it is the ``pure"  preferential attachment mechanism (the triangle-growth function) and Fibonacci series (the star-growth function and the rectangle-growth function) that together make graph-model $N(t)$ not only scale-free and small-world, but also the typical degree distribution parameter $\gamma=3$. In addition, there are differences between them. In BA model, the probability $P_{v}(i)$ of any vertex $v$ obtaining new edges only depends on $v$ current status, namely its degree $k_{v}(i)$, with ¡°pure¡± preferential attachment mechanism at time step $i$. In other words, this previous degree sequences $k_{v}(j)$ ($j<i$) of vertex $v$ make no influence on probability $P_{v}(i)$. Hence, some researchers propose and make use of Markov model to study the development of complex networks \cite{B-C-2018, F-S-2018}. However, in our model, the probability $P_{v}(i)$ not only rely on vertex $v$ current status $k_{v}(i)$ but also is related to the processing status $k_{v}(i-1)$ with respect to the properties of Fibonacci series. Our result is only a tip of the iceberg, a large amount of work need to be done in the future.

\section{Application to Brouwer's fixed point theorem}

Before starting to enumerate the number of spanning trees over graph-model $N(t)$, we have to introduce an application to the
well-known Brouwer's fixed point theorem.

\textbf{Lemma 1} If two conditions, $r\neq0$ and $ps-qr\neq0$, hold on the formula $a_{n+1}=\frac{p\cdot a_{n}+q}{r\cdot a_{n}+s}$, then the solution of $a_{n}$ can be calculated, as follows
\begin{equation}\label{eqa:mf-5-1}
\left\{\begin{split}&a_{n}=\frac{(a_{1}\mu-\lambda\mu)\left(\frac{p-\lambda r}{p-\mu r}\right)^{n-1}-(a_{1}\mu-\lambda\mu)}{(a_{1}-\lambda)\left(\frac{p-\lambda r}{p-\mu r}\right)^{n-1}-(a_{1}-\mu)},~  \lambda\neq \mu\\
&a_{n}=\frac{a_{1}-\lambda}{\left(a_{1}\frac{2r}{p+s}-\lambda\frac{2r}{p+s}\right)n+1-a_{1}\frac{2r}{p+s}+\lambda},~  \lambda= \mu
\end{split}
\right.
\end{equation}
where parameters $\lambda$ and $\mu$ satisfy an equality about variable $x$, $rx^{2}+(s-p)x-q=0$. For further understanding lemma 1, reader is encouraged to see  \cite{F-J-Y-2018}.

\section{Enumeration of spanning trees on  model $N(t)$ }

As known, computing the number of spanning trees of any finite graph had been theoretically shown by the well-known Kirchhoff's matrix-tree theorem, that is to say, the value of spanning tree number is equivalent to the product of all nonzero eigenvalues of the Laplacian matrix of the graph. When we sometime consider an enormous network model
with hundreds and thousands of vertices and edges, such a theoretical accomplishment will be difficultly implemented only because of the computing difficulty. But spanning trees number is always treated as an important invariant correlated to several kinds of dynamical functions on models, for instance reliability \cite{G-J-S-2003}, synchronization capability \cite{N-T-A-E-2006}, random walks \cite{P-M-2000}-\cite{Z-Z-2013}, to name just a few. Hence, lots of works about computing the number of spanning trees of special network models need to be done in the future. In addition, a group of elements with interesting topological structure properties in spanning tree set have already gotten much consideration from different disciplines, from discrete mathematics (as an active branch of mathematics) to statistic physics, chemistry and biology, again to theoretical computer sciences. These uncommon elements have all a type of special structure, the leaf number being maximum, which are called maximum leaves spanning tree, shorted as MLS-tree. In particular, such topological structure on spanning trees performs a significant figure to design and maintain networks which are in an open environment. It seems to be a simple example serving as to clarify the realistic application that most of people always acquire useful information by visiting specific web-site in which each smartphone, laptop or desktop can be abstractly seen as a pendent vertex (or a leaf) on the whole network built by millions of smart-devices.

Similar to discussions in subsection 4.3, we divide complex computational procedure about spanning tree number into two parts, following

\textbf{Lemma 2} For each generated triangle on graph-model $N(t)$, the corresponding solution of the total number of spanning trees is

\begin{equation}\label{eqa:mf-6-1}
S_{1}(t)=2^{3^{t-1}}\times\left(\frac{3}{2}\right)^{t}\prod_{i=0}^{t-2}2^{3^{i}}\left(\frac{3}{2}\right)^{3^{i}(t-i-1)}.
\end{equation}

\textbf{Lemma 3} For each rectangle generated by rectangle-growth function on graph-model $N(t)$, the exact solution of the total number of spanning trees is

\begin{equation}\label{eqa:mf-6-2}
Q(t)=2^{1+2\sum_{i=0}^{t-1}3^{i}}\left(\frac{3}{2}\right)^{2\sum_{i=0}^{t-2}3^{i}(t-1-i)}\left[\left(\frac{3}{2}\right)^{t}+\left(\frac{3}{2}\right)^{2t}\right].
\end{equation}

We here omit some analytical derivation on computing $S_{1}(t)$ and $Q(t)$. For further read, seeing \cite{F-M-20181}. With respect to both the growth mechanism of $N(t)$ and the two values of $S_{1}(t)$ and $Q(t)$, we may gain

\textbf{Proposition 6} The solution of the total number of spanning trees of graph-model $N(t)$ is

\begin{equation}\label{eqa:mf-6-3}
\mathbb{S}(t)=4^{\Theta(t)}\times Q(t)\times Q^{4}(t-2)\prod_{i=1}^{t-3}Q(i)^{\Theta(t-i)}
\end{equation}
in which superscript $\Theta(t)$ has same meanings as one at table 3.

Next, we will illustrate this contribution from three different types of operations, triangle-growth function, star-growth function and rectangle-growth function, respectively, to the construction of spanning trees with maximum leaves, shown in Fig.3. The notation $\Psi(t)$ denotes the number of leaves on any spanning tree with maximum leaves. Let us organize the procession by adopting an increasing order of computational complexity, as follows

\begin{figure}
\centering
  \includegraphics[height=4cm]{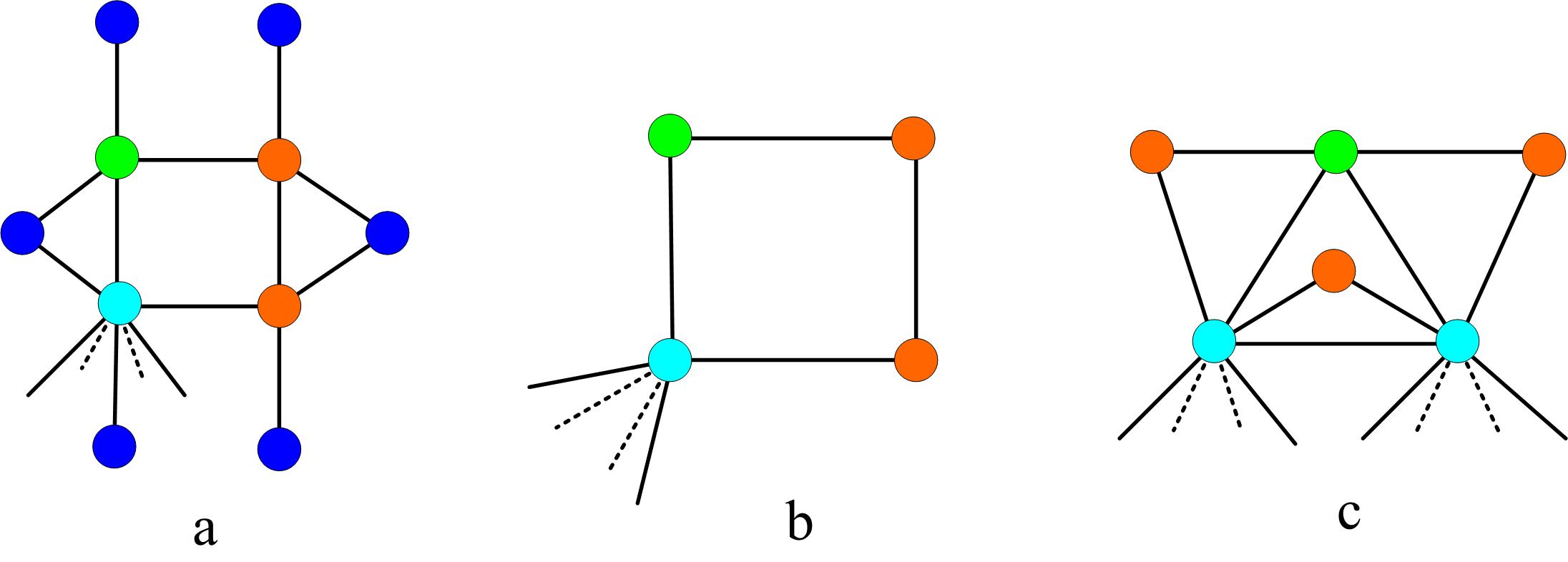}\\
{\small Fig.3. The diagram of this contribution from three different types of operations to the construction of spanning trees with maximum leaves on model $N(t)$, in which indigo vertex is followed by green, green by orange and orange by blue in the evolution of model $N(t)$. }
\end{figure}

\textbf{Case 1} It may be obviously seen that each new vertex added by star-growth function must be a leaf of arbitrary spanning tree with maximum leaves, and then we denote this part leaf number by $\Psi_{1}(t)$. After applying this operation, it is impossible that each vertex seated at any rectangle on $N(t)$ is a leaf of arbitrary maximum leaves spanning tree, shown in Fig.3(a).

\textbf{Case 2} For two new vertices generated by rectangle-growth function, it is possible both vertices all are leaves of arbitrary spanning tree with maximum leaves. In addition, any one of which and one younger of two remained vertices on this new resulting rectangle can also be leaves of arbitrary spanning tree with maximum leaves. This part leaf number is indicated by notation $\Psi_{2}(t)$. Readers may easily understand case 2 by seeing Fig.3(b).

\textbf{Case 3} By analogy analysis as the above two cases, three new vertices presented by triangle-growth function
will all be leaves of arbitrary spanning tree with maximum leaves. Except for these new vertices, there is a portion of vertices being leaves of arbitrary spanning tree with maximum leaves. For the second part, we find that only vertices added dy triangle-growth function at time step $t-1$ may be leaves of arbitrary spanning tree with maximum leaves and those remained vertices on each triangle are not, seeing Fig.3(c). If no, suppose that a vertex added by triangle-growth function at time step $t-2$ is a leaf, and then this manner will generate a new tree at the expense of decreasing at least a leaf in comparison with our desirable trees. Considering the discussion related to other vertices is similar to one at time step $t-2$, we omit the calculation. The sum of both classes of leaf number is represented by notation $\Psi_{3}(t)$.

Take into account cases 1-3, we have

\textbf{Proposition 7} The leaf number of arbitrary maximum leaves spanning tree of graph-model $N(t)$ is

\begin{equation}\label{eqa:mf-6-4}
\Psi(t)=\Psi_{1}(t)+\Psi_{2}(t)+\Psi_{3}(t)=17\times4^{t-2}+\sum_{i=3}^{t-1}3^{t-1-i}\left(2\times\frac{4^{i-1}-4}{3}\right).
\end{equation}

\textbf{Proposition 8} The solution of the total number of maximum leaves spanning trees of graph-model $N(t)$ is

\begin{equation}\label{eqa:mf-6-5}
\mathcal{S}(t)=3^{\frac{4^{t-1}-4}{3}}\times2^{\sum_{i=3}^{t-1}3^{t-1-i}\left(2\times\frac{4^{i-1}-4}{3}\right)}\times\mathbb{S}(t-2).
\end{equation}

Generally speaking, the complexity of calculating the total number of spanning trees is almost equivalent to enumerating all spanning trees with maximum leaves on a larger graph-model. As one of our innovation viewpoints, the relationship between the total number of spanning trees and the number of all spanning trees with maximum leaves is built. The reason for this relation is maybe three given operations, which together determine our graph-model $N(t)$ has scale-free feature and small-world property. In other words, we provide a question: is there a brief connection as Eq.(\ref{eqa:mf-6-5}) between the total number of spanning trees and the number of all maximum leaves spanning tree on most of deterministic models possessing scale-free feature and small-world property. If so, it should be available and meaningful to seek for such a bridge through making use of our thoughts and discussions here.

\section{Conclusion and problem}

In this paper, we propose a new graph-model to study complex network and discuss several typical parameters seen in a wide variety of published literatures, including 1) average degree, 2) degree distribution, 3) clustering coefficient, 4) diameter and $APL$ as well as 5) spanning trees number and the number of spanning trees with maximum leaves. Eq.(\ref{eqa:mf-4-1-1}) represents our graph-model is sparse. Graph-model $N(t)$ turns out to be scale-free in Eq.(\ref{eqa:mf-4-2-1}). Based on Eq.(\ref{eqa:mf-4-3-6}), Eq.(\ref{eqa:mf-4-4-1}) and Eq.(\ref{eqa:mf-4-4-2}), we show small-world property appears on graph-model $N(t)$. Eq.(\ref{eqa:mf-6-3}) and  Eq.(\ref{eqa:mf-6-5}) say the spanning trees number and the solution to the total number of maximum leaves spanning trees, respectively. Despite the importance of discussing the above statements, we here should again claim our innovation viewpoints, as follows

(a) Find one relation between power-law and zipf-law on deterministic graph-model. As seen, the study of power-law distribution and zipf-law distribution is an active branch in which there is amounts of current research interest and consideration discussions. Though a lager number of research papers have be published to consider connection between both, no any accepted result is reported. The statements and explanations presented here certainly provide some insights, but there is still much work to be done to make us better understand the inherent mechanism between power-law and zipf-law both experimentally and theoretically.

(b) Build a bridge connecting Fibonacci series and ``pure" preferential attachment mechanism. This is an exciting outcome which reveals mathematical beauty. As mentioned at the end of section 4, our result has more a large amount of technological practice values and potential theoretical meanings. It seems to have lots of intriguing discoveries behind topological structure of complex networks before we can really understand the dynamical processes driving these complex systems evolution.

(c) Generate a equation relating spanning trees number and the total number of maximum leaves spanning trees on a graph-model. The above two structural parameters not just have been studied deeply at the theory level, such as combinatorial mathematics, statistic physics, etc., but also are widely used at the technical level, including reliability of fault-tolerance to random faults and of intrusion-tolerance to selectively remove attacks, synchronization capability and diffusion properties of networks, and so on. In order to better capture relationship between the two, we have a try to quantitatively determine this problem on a deterministic model with scale-free feature and small-world property. Without doubt it is a long path that we can ultimately unveil this problem from the practical point of view.

Taking into account the above mentioned descriptions and our next research directions, there are three corresponding problems, as follow

(a') Is there such a simple relation described in Eq.(\ref{eqa:mf-4-2-5}) in all complex systems which obey other power-law or zipf-law? If so, it should be noteworthy, one of many reasons is the current password distribution seems to obey the zipf-law. And then, arbitrary app using password as authentication can be studied deeply at some case, for example, the research about digital virtual asset security threat perception based on password.

(b') As seen, the length 2 Fibonacci series seems to have the same function on growth of network models as the ``pure" preferential attachment mechanism. Does the Fibonacci series of any length have such a performance?

(c') Does any deterministic network model having scale-free feature and small-world property possess such an equation similar to Eq.(\ref{eqa:mf-6-5})?

In a word, there is no doubt that there will be many interesting discoveries and formidable challenge still waiting to be done in the days to come.

\textbf{Acknowledgment.} The research was supported by the National Key Research and Development Plan under grants 2016YFB0800603 and 2017YFB1200704, and the National Natural Science Foundation of China under grant No. 61662066.

\vskip 1cm

\textbf{Appendix. Complementary material}

Two widely discussed deterministic models should serve as theoretical material to support our assertion: all deterministic scale-free models follow the expression $\frac{f^{\lambda}_{r_{i}}}{P_{cum}(k\geq k_{r_{i}})}\propto O(1)$, where exponent $\lambda$ is a constant as $t\rightarrow \infty$. Here, in order to clarify our purpose, we just collect degree sequence on both network models whose other topological properties can be easily read in variety of published literatures, such as \cite{J-S-A-J-2005,J-H-G-2009}

\textbf{A. Apollonian network model}

In reference \cite{J-S-A-J-2005}, the Apollonian network model has scale-free feature, $P_{cum}(k\geq k_{r_{i}})=k_{r_{i}}^{1-\gamma}$ ($\gamma=1+\frac{\ln 3}{\ln 2}$). The below table consists of rank, vertex-number and vertex-degree.

\small \textbf{Table 6.}
\begin{center}
\begin{tabular}{c|c|ccccc}
\hline
   $rank$& $k$& $n$ \\
\hline
  $1$& $3\times 2^{t}$& $1$ \\
\hline
 $2$& $3\times 2^{t-1}$& $3$\\
\hline
 $3$& $3\times 2^{t-2}$& $3^{2}$ \\
\hline
 $4$& $3\times 2^{t-3}$& $3^{3}$ \\
\hline
  $...$& $...$& $...$\\
\hline
 $t_{i}+1(=r_{i})$&   $3\times 2^{t-t_{i}}$& $3^{t_{i}}$\\
\hline
  $...$&  $...$& $...$& \\
\hline
 $t$&  $3\times 2$& $3^{t-1}$\\
\hline
 $t+1$&  $3$& $3^{t}$\\
\hline
\end{tabular}
\end{center}

$$f_{r_{i}}=\frac{\sum_{r\geq r_{i}}n_{r}}{2|E(t)|}\sim \frac{3^{t_{i}+2}-3\times 2^{t_{i}+1}}{3^{t+2}-3\times 2^{t+1}}\sim k_{r_{i}}^{1-\gamma}$$

where $\gamma$ is equal to $1+\frac{\ln 3}{\ln 2}$, showing that the Apollonian network model satisfies our assertion.

\textbf{B. Sierpinski network model}

In reference \cite{J-H-G-2009}, the Sierpinski network model possesses scale-free feature and has the different  degree-distribution exponent from the Apollonian network model, $P_{cum}(k\geq k_{r_{i}})=k_{r_{i}}^{1-\gamma}$ ($\gamma=2+\frac{\ln 2}{\ln 3}$). The table 7 lists out rank, vertex-number and vertex-degree.

\small \textbf{Table 7.}
\begin{center}
\begin{tabular}{c|c|ccccc}
\hline
   $rank$& $k$& $n$ \\
\hline
  $1$& $3^{t}+1$& $6$ \\
\hline
 $2$& $3^{t-1}+1$& $3\times6$\\
\hline
 $3$& $3^{t-2}+1$& $3\times6^{2}$ \\
\hline
 $4$& $3^{t-3}+1$& $3\times6^{3}$ \\
\hline
  $...$& $...$& $...$\\
\hline
 $t_{i}(=r_{i})$&   $3^{t-t_{i}+1}+1$& $3\times6^{t_{i}-1}$\\
\hline
  $...$&  $...$& $...$& \\
\hline
 $t$&  $3+1$& $3\times6^{t-1}$\\
\hline
\end{tabular}
\end{center}

$$f_{r_{i}}=\frac{\sum_{r\geq r_{i}}n_{r}}{2|E(t)|}\sim \frac{3\times6^{t_{i}}+3/5(6^{t_{i}-1}-1)+6}{3\times6^{t}+3/5(6^{t-1}-1)+6}\sim k_{r_{i}}^{1-\gamma}$$

here $\gamma$ is equivalent to $2+\frac{\ln 2}{\ln 3}$, supporting again our assertion.

\end{document}